\documentclass[letterpaper, twocolumn, pra]{revtex4}
\usepackage{graphicx, amsmath, amssymb, bm, titlesec, mathtools}
\DeclareMathOperator{\sinc}{sinc}

\begin{document}

\title{Controlling propagation of spatial coherence for enhanced imaging through scattering media}

\author{Abhinandan Bhattacharjee, Shaurya Aarav, Harshawardhan Wanare, and Anand K. Jha}

\email{akjha9@gmail.com}

\affiliation{Department of Physics, Indian Institute of
Technology Kanpur, Kanpur UP 208016, India}

\date{\today}

\begin{abstract}
It is known that a spatially partially coherent light field produces better imaging contrast compared to a spatially coherent field and that the contrast increases as the spatial coherence length of the field becomes smaller. The transverse spatial coherence length of most spatially partially coherent fields increases upon propagation. As a result, the field produces progressively decreasing image quality at subsequent transverse planes. By controlling the propagation of spatial coherence, we demonstrate enhanced image quality at different transverse planes along the propagation direction through a scattering medium. Using a source with propagation-invariant spatial coherence function, we report experimental observations of imaging different transverse planes with equal contrast over a significant distance. Furthermore, we generate a spatially partially coherent source that can be tailored to have minimum-possible transverse coherence area at the plane of the object to be imaged, and using this source, we demonstrate imaging spatially separated transverse planes with maximum possible image contrast. 
\end{abstract}

\maketitle

\section{Introduction}
Imaging through scattering   media is a very important area of research due to its implications for a wide range of real-world applications. For example, imaging of objects at different transverse planes through atmospheric fog is inevitable in daily life scenarios such as railways, defence, and road transports. Imaging through scattering media has been an important research problem since 1960s \cite{goodman1966apl, kogelnik1968josa}, and even today this a very active area of research \cite{kang2015natphot, katz2014natphot, bertolotti2012nature, liba2017natcomm}. The difficulties in imaging through scattering   media arise due to the inhomogeneities in such media which introduce random phase variations at different spatial locations in the light field passing through it. If the light field is spatially coherent, these random phase variations result in a random interference pattern known as the speckle pattern \cite{goodman2007speckle}. As a consequence, what gets recorded is the image of the object superimposed with the speckle pattern at the imaging plane. Thus the recorded image gets corrupted and the image quality gets severely affected  \cite{ntziachristos2010natmed}.

Over the years, several imaging techniques have been developed for addressing the difficulties caused by speckle effects in scattering media. These techniques can be categorized into two sets. The first set of techniques is based on using spatially completely coherent light sources such as lasers for illumination.  In this set of techniques, one tries to minimize the speckle effects either by imaging with ballistic photons \cite{huang1991science, liba2017natcomm, velten2012natcomm, leith1992josaa} or by descrambling the phase of the scattered light field using a hologram or a spatial light modulator (SLM)  \cite{kogelnik1968josa, harm2014optexp, vellekoop2007optlett, mosk2012natphot}. The other set of techniques for imaging through scattering   media is based on using spatially partially coherent light sources. In this set of techniques, the speckle effect gets reduced as the spatial coherence length of the field becomes smaller.
There are several different approaches to generating spatially partially coherent light fields. The most common approach involves introducing randomness in a spatially coherent laser field by using either an acousto-optical cell \cite{ohtsuka1980optcomm}, a rotating ground glass plate \cite{chen2014pra, wang2014optexp}, or an SLM \cite{basu2014optexp, ostrovsky2009optexp}. A more recent approach involves using random lasers \cite{redding2012natphot,redding2015pnas} with small spatial coherence lengths. The other approach is to use light-emitting diodes (LEDs) or thermal sources \cite{carter1977josa, tziraki2000apb}, which are spatially completely incoherent primary light sources. Although the techniques based on spatially completely coherent sources such as lasers are useful for some applications requiring intense illumination, they still have only limited applicability in full-field imaging due to speckle effects. As a result, the techniques based on using spatially partially coherent sources are being preferred for imaging two-dimensional objects in scattering  media \cite{redding2012natphot, redding2015pnas, redding2015optlett, knitter2016optica, zheng2016optexp, farrokhi2017scirpt}.
However, the spatial coherence length of most partially coherent sources increases upon propagation causing speckle effects to become progressively pronounced. Therefore, such sources become unsuitable for imaging spatially separated transverse planes along the propagation direction. 

In this article, we demonstrate that the above issue can be overcome through controlling the propagation of spatial coherence of partially coherent sources. First of all, we report a proof-of-principle experimental demonstration of imaging different transverse planes with equal contrast over a distance of 40 cm along the propagation direction. This is achieved using a recently demonstrated source in which the spatial coherence is controlled in a manner that the spatial coherence function remains propagation-invariant \cite{aarav2017pra}. Next, we demonstrate a source in which the propagation of spatial coherence is controlled in order to yeild the minimum-possible transverse coherence area at the plane of the object to be imaged. Using such a partially coherent source, we demonstrate imaging different transverse planes along the propagation direction with the maximum possible contrast.

\begin{figure}[t!]
\centering
`\includegraphics{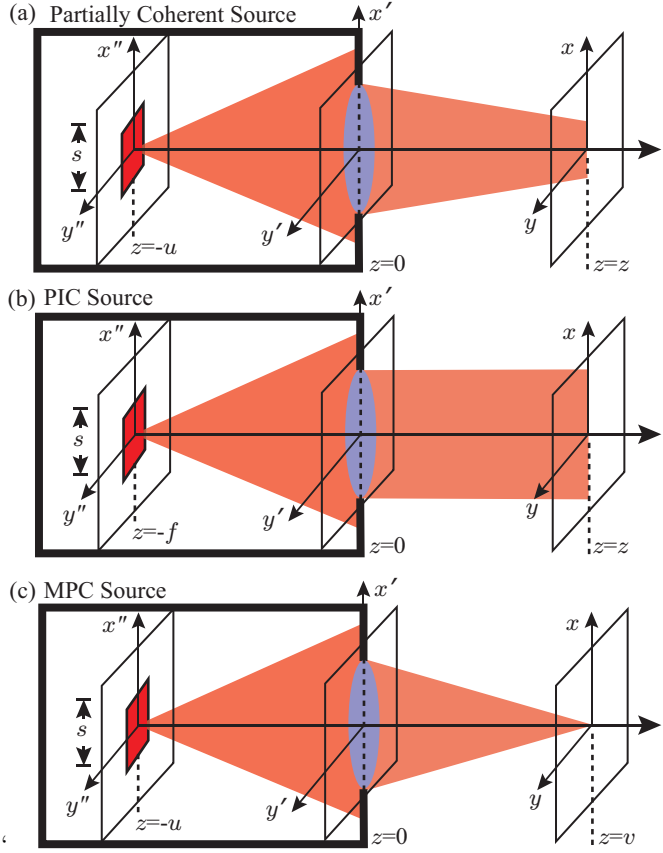}
\caption{Schematic illustration of (a) a sptially partially coherent source, (b) a propagation-invariant coherence (PIC) source, and (c) a minimum-possible coherence (MPC) source}\label{fig1}
\end{figure}

\section{Controlling the propagation of spatial coherence}

Figure \ref{fig1}(a) shows a generic spatially partially coherent source, in which a planar monochromatic spatially completely incoherent source is kept at a distance $u$ behind a lens located at $z=0$. We represent the transverse spatial location at $z=-u$ by $\bm{\rho''}\equiv (x'', y'')$ and that at $z=0$ and $z=z$  by $\bm\rho'\equiv (x', y')$ and $\bm\rho\equiv (x, y)$, respectively. The primary incoherent source along with the lens constitute our spatially partially coherent source. Since our primary source is spatially completely incoherent, the cross spectral density of the field at $z=-u$ is given by  
\begin{align}
W(\bm\rho_1'',\bm\rho_2'',z=-u)=NI_s(\bm\rho_1'',z=-u)\delta(\bm\rho_1''-\bm\rho_2''),\label{incoherent}
\end{align}
where $I_s(\bm\rho_1'', z=-u)$ is the intensity of the primary source at $z=-u$ and is  given by $I_s(\bm\rho'',z=-u)=A$, if $-s/2<x''<s/2$ and $-s/2<y''<s/2$, else 0 with $A$ being a constant. $N=\tfrac{\lambda_0^2}{\pi}$ (see Ref~\cite{goodman2015stat}, section 5.5.4), where $\lambda_0$ is the central wavelength. Following Section 4.4.3 of \cite{mandel1995coherence}, we write the cross-spectral density function $W(\bm\rho_1,\bm\rho_2,z)$ of the field at $z$ in terms of the cross-spectral density function $W(\bm{\rho_1'},\bm{\rho_2'},z=0)$ of the field at $z=0$ right after the converging lens as  
\begin{multline}
W(\bm\rho_1,\bm\rho_2,z)=\tfrac{1}{z^2}e^{\tfrac{ik_0}{2z}(\rho_2^2-\rho_1^2)} \iint W(\bm{\rho'_1},\bm{\rho'_2},z=0) \\ \times e^{\tfrac{ik_0}{2z}(\rho'^2_2-\rho'^2_1)}
e^{\tfrac{-ik_0}{z}(\bm{\rho_2}\cdot\bm{\rho'_2}-\bm{\rho_1}\cdot\bm{\rho'_1})} d\bm{\rho'_1}d\bm{\rho'_2}.
\label{cs-density-pi-3}
\end{multline}
Here $k_0=\omega_0/c$ with $\omega_0$ being the central frequency of the field, and $\rho_1=|\bm\rho_1|$, $\rho_2=|\bm\rho_2|$, etc. The cross spectral density $W(\bm{\rho'_1},\bm{\rho'_2},z=0)$ after the lens can be calculated by propagating the cross-spectral density at $z=-u$ until $z=0$ before the lens and then propagating it through the lens. This way we obtain
\begin{multline}
W(\bm{\rho'_1},\bm{\rho'_2},z=0)=\frac{1}{u^2}e^{\tfrac{ik_0}{2u}(\rho'^2_2-\rho'^2_1) } T^*(\bm{\rho_1'})T(\bm{\rho_2'}) \\ \times \iint  W(\bm{\rho_1''},\bm{\rho_2''},z=-u)  e^{\tfrac{ik_0}{2u}(\rho_2''^2-\rho_1''^2)}
\\ \times e^{\tfrac{-ik_0}{u}(\bm{\rho'_2}\cdot\bm{\rho_2''}-\bm{\rho'_1}\cdot\bm{\rho_1''})}  d\bm{\rho_1''}d\bm{\rho_2''}.\label{cs-density}
\end{multline}
Here $T(\bm\rho)$ is the amplitude transmittance function of the lens and is given by $T(\bm\rho)=\exp(-ik_0\rho^2/2f)$, where $f$ is the focal length of the lens. Substituting the expressions for the amplitude transmission function and also that of the cross-spectral density function $W(\bm{\rho_1''},\bm{\rho_2''},z=-u)$ of Eq.~\ref{incoherent} into Eq.~(\ref{cs-density}), evaluating the $\bm\rho_2''$ integral, and replacing $\bm\rho_1''$ by $\bm\rho''$, we can write Eq.~(\ref{cs-density}) as
\begin{multline}
W(\bm{\rho'_1},\bm{\rho'_2},z=0)=\frac{AN}{u^2}e^{\tfrac{ik_0}{2}(\rho'^2_2-\rho'^2_1)(\tfrac{1}{u}-\tfrac{1}{f}) }\\ \times \int 
e^{\tfrac{-ik_0}{u}(\bm{\rho'_2}-\bm{\rho'_1})\cdot\bm{\rho''}} d\bm{\rho''}. \label{cs-density-pi-2}
\end{multline}
Now, substituting Eq.~(\ref{cs-density-pi-2}) into Eq.~(\ref{cs-density-pi-3}), we obtain the cross-spectral density function $W(\bm\rho_1,\bm\rho_2,z)$ at $z$:
\begin{multline}
W(\bm\rho_1,\bm\rho_2,z)=\frac{AN}{u^2 z^2} e^{\tfrac{ik_0}{2z}(\rho_2^2-\rho_1^2)} \iiint e^{-\tfrac{ik_0}{u} (\bm{\rho'_2}-\bm{\rho'_1})\cdot\bm{\rho''} }  \\ \times  e^{-\tfrac{ik_0}{2\Delta(z)} (\rho'^2_2-\rho'^2_1)} 
e^{-\tfrac{ik_0}{z}(\bm{\rho_2}\cdot\bm{\rho'_2}-\bm{\rho_1}\cdot\bm{\rho'_1})} d{\bm\rho''} d\bm{\rho'_1}d\bm{\rho'_2}
\label{cs-density-pi-4}
\end{multline}
where $\frac{1}{\Delta(z)}=\frac{1}{f}-\frac{1}{u}-\frac{1}{z}$. This is the general expression for $W(\bm\rho_1,\bm\rho_2,z)$. We note that the lens is symmetric with respect to $x''$ and $y''$. Thus, $W(\bm\rho_1,\bm\rho_2,z)$ can be written as $W(\bm\rho_1,\bm\rho_2,z)=W(x_1,x_2,z)W(y_1,y_2,z)$. For conceptual clarity, we numerically solve only the $x$-integral which is given by
\begin{multline}
W(x_1,x_2,z)=\frac{\sqrt{AN}}{u z} e^{\tfrac{ik_0}{2z}(x_2^2-x_1^2)} \iint_{-D/2}^{D/2} \int_{-s/2}^{s/2}   \\ \times  e^{-\tfrac{ik_0}{u} (x'_2-x'_1)x'' } e^{-\tfrac{ik_0}{2\Delta(z)} (x'^2_2-x'^2_1)}  
e^{-\tfrac{ik_0}{z}(x_2 x'_2-x_1 x'_1)} 
\\ \times d x'' dx'_1 d x'_2,
\label{cs-density-pi-5}
\end{multline}
\begin{figure}[t!]
\centering
\includegraphics{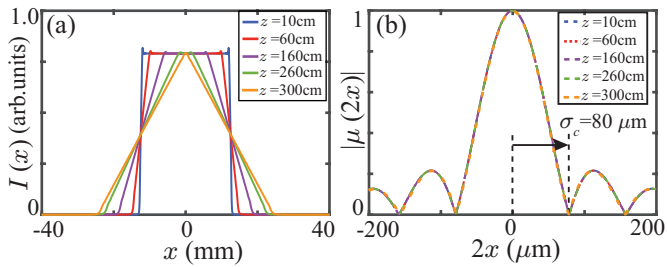}[t!]
\caption{Plots of (a) intensity $I(x)$ and (b) the degree of coherence $|\mu(2x)|$ of the PIC source for various $z$ values.}\label{fig2_mod}
\end{figure}
The integral over $x''$ needs to be evaluated over the source size, that is, from $-s/2$ to $s/2$ while the integrals over $x_1'$ and $x_2'$ needs to be evaluated over the size of the lens which we take to be $D$. We are interested in the cross-spectral density function that is symmetric about the $z$-axis. Thus, by substituting $x_1=x$ and $x_2=-x$ and then evaluating the $x''$ integral, we obtain the following expression for the symmetric cross-spectral density function $W(x,-x,z)$ and the corresponding intensity $I(x)=W(x,x,z)$:
\begin{align}
&W(x,-x,z) = \frac{\sqrt{AN}s}{u z}\iint_{-D/2}^{D/2} \sinc\left\{\tfrac{k_0s}{2u}( x'_2-x'_1)\right\} \notag \\ & \qquad\qquad\quad \times e^{-\tfrac{ik_0}{2\Delta(z)}(x'^2_2-x'^2_1)}  
e^{\tfrac{ik_0}{z}(x'_2+x'_1)x} d{x'_1}d{x'_2}. \label{cs-density-fininte-6} \\
&I(x,z)= \frac{\sqrt{AN}s}{u z}\iint_{-D/2}^{D/2} \sinc\left\{\tfrac{k_0s}{2u}( x'_2-x'_1)\right\} \notag \\ & \qquad\qquad\quad \times e^{-\tfrac{ik_0}{2\Delta(z)}(x'^2_2-x'^2_1)} 
e^{-\tfrac{ik_0}{z}(x'_2-x'_1)x} d{x'_1}d{x'_2}.\label{intensity-7}
\end{align}
The degree of coherence function $|\mu(x, -x, z)|$ is given by
\begin{align}
|\mu(x, -x, z)|=|\mu(2x, z)|=W(x, -x, z)/ I(x, z) \label{degree of coh}
\end{align}
We take the half width $\sigma_c$ of this function as the transverse spatial coherence length.  We next evaluate $\mu(2x,z)$ and $I(x,z)$ for two special cases.

\subsection{Propagation-Invariant Coherence source $(u=f)$ }

We consider the situation in which $u=f$, that is, when the primary incoherent source is kept at the back focal plane of the lens. Figure \ref{fig1}(b) shows the configuration of the source in this case. It has been shown in Ref.~\cite{aarav2017pra} that when the aperture size of the lens is infinite, the degree of coherence function $|\mu(2x,z)|$ and the intensity $I(x,z)$ become independent of $z$. Even when the aperture size is finite the degree of coherence function remains $z$-independent up to the distance given by $z_{\rm max}=Df/s$. Therefore, such sources are referred to as the propagation-invariant coherence (PIC) source. We numerically evaluate Eqs.~(\ref{intensity-7}) and (\ref{degree of coh}) for $D=2.5$ cm, $f=10$ cm $s=0.8$ mm, and  plot $I(x,z)$  and $|\mu(2x, z)|$ in Fig.~\ref{fig2_mod}(a) and Fig.~\ref{fig2_mod}(b), respectively, for various values of $z$. We find that while the intensity profile of the source starts to broaden as a function of $z$, the degree of coherence function remains independent of $z$, that is, it remains propagation invariant up to 300 cm. Taking the distance to the first zero of $|\mu(2x,z)|$ function as $\sigma_c$, we find it to be about 80 $\mu$m. 
\begin{figure}[t!]
\centering
\includegraphics{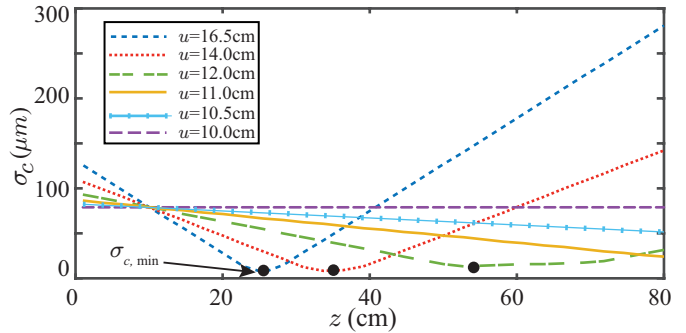}
\caption{Plots of transverse spatial coherence length $\sigma_c$ of the MPC source as a function of $z$ for various values of $u$. The minimum $\sigma_{c, \rm min}$ appears near $z=v$, where $v$ is the image plane of the primary incoherent source.}\label{MPC}
\end{figure}

\subsection{Minimum-Possible Coherence source $(u>f)$}
Next, we consider the situation in which $u>f$ (see Figure \ref{fig1}(c)). Using Eq.~(\ref{degree of coh}), we numerically evaluate $|\mu(2x,z)|$ as a function of $x$, and taking the distance to its first zero to be $\sigma_c$, we calculate and plot $\sigma_c$ as a function of $z$ for various values of $u$ (see Fig.~\ref{MPC}). For a given $u$, $\sigma_c$ decreases with $z$ and reaches its minimum possible value $\sigma_{c,{\rm min}}$ near $z=v$, where $v$ is the image distance of our primary source. As $u$ is decreased, $v$ increases and therefore the $z$ value at which $\sigma_{c,{\rm min}}$ appears shifts to the right with $\sigma_{c,{\rm min}}$ remaining almost constant. Thus, we refer to this source as minimum-possible coherence (MPC) source. It can be used for imaging two-dimensional objects kept at $z=v$ with maximum possible imaging contrast. Furthermore, within the $z$-range over which $\sigma_{c,{\rm min}}$ remains almost constant, a two-dimensional object could be placed at any $z$ and be imaged with maximum possible contrast by adjusting $u$ in a way that $\sigma_{c,{\rm min}}$ appears at the given $z$. 
For $D=2.5$ cm, $f=10$ cm $s=0.8$ mm, $\sigma_{c,{\rm min}}$ changes from 6.5 $\mu$m to about 8.5 $\mu$m from $z=25$ cm to $z=35$ cm. We note that when $u$ approaches $f$, the MPC source becomes the PIC source.

\section{Experimental results: enhanced imaging contrast through scattering media}

We next present our experimental results demonstrating how PIC and MPC sources can be used for imaging different transverse planes with enhanced imaging contrast through scattering media. In our experiments, we use lab-synthesized ground glass plates and stack together varying number of them in order to get scattering media of different scattering strengths. We characterize the strength of thus-constructed scattering media in the following manner. We make a laser beam pass through the scattering medium whose strength we need to measure. We record the intensity of a small central portion of the laser beam on a $50\times 50$-pixel area of the CCD camera, kept at a distance of $30$ cm from the scattering medium.  The measured intensity in the presence and in the absence of the scattering medium is called $I$ and $I_0$, respectively. For our scattering media, the material absorption is negligible; so any drop in the recorded laser intensity in the presensce of a scattering medium is solely due to scattering. Therefore,  we take the ratio $I_0/I$ of the two intensities as the scattering strength  of the medium and write it as  $\alpha=\tfrac{I_0}{I}$. The quantity  can be shown to be related to the scattering coefficient $\mu_s$ as $\alpha=e^{\mu_s d}$, where $d$ is the thickness of the scattering medium \cite{wang2012biomedical}. We note that in our experiments we use scattering media of varying scattering coefficient $\mu_s$ and thickness $d$. Therefore, for characterizing the strength of our scattering media, $\alpha=e^{\mu_s d}$ is a more pertinent quantity instead of $\mu_s$. Larger values of $\alpha$ represent increased scattering strength, with $\alpha=1$ representing no scattering.

\subsection{Imaging with a propagation-invariant source in transmitting configuration}

\begin{figure}[t!]
\centering
\includegraphics{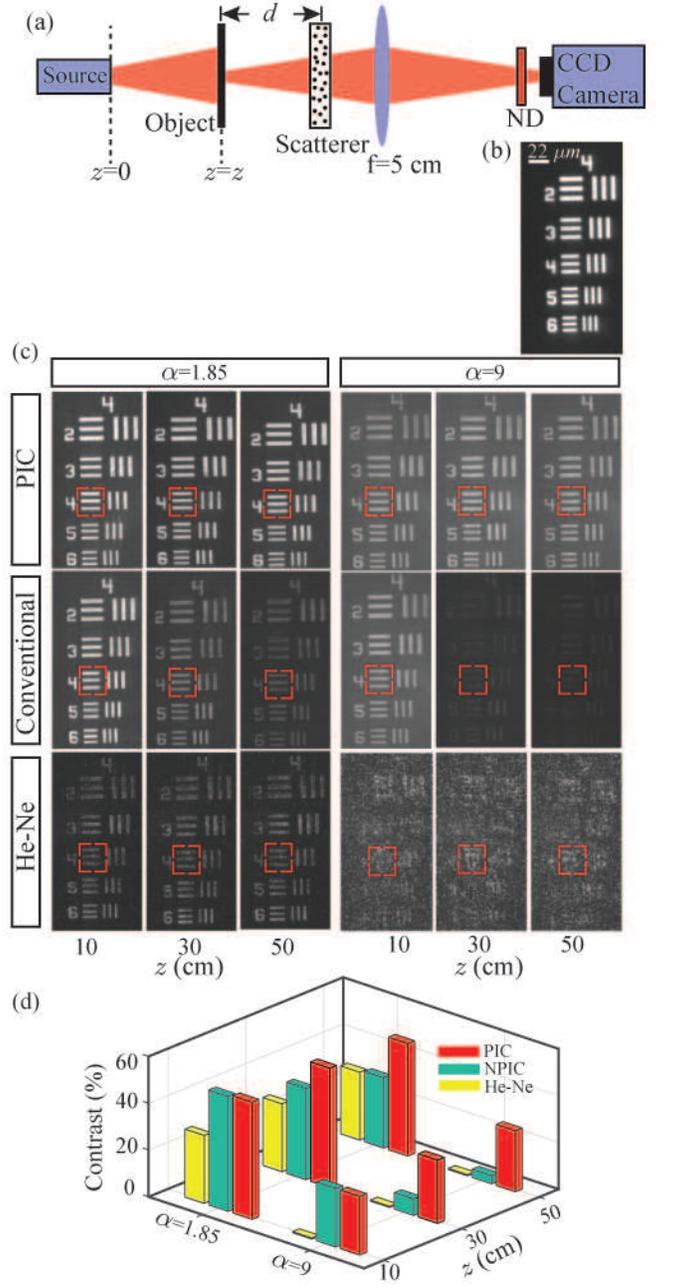}
\caption{(a) Schematic of the setup for imaging in transmitting configuration. (b) Image of the object in the absence of scattering. (c) Images of the object and (d) imaging contrast obtained with the three different sources at different $z$ and $\alpha$ values. 
}\label{fig2}
\end{figure}

First of all, we present our experimental results of imaging through scattering medium in transmitting configuration with a PIC source and compare its performance with that of a spatially coherent source and a conventional partially coherent source wherein the transverse coherence width increases with propagation. Figure \ref{fig2}(a) shows the schematic diagram of the experimental setup. A source kept at $z=0$ illuminates a transmission object kept at $z=z$. We use a 632-nm, 5-mW HeNe laser having a Gaussian beam profile as a spatially coherent source, while for the conventional source we use an LED. We consider the same LED as the primary source in the configuration of PIC source, as shown in Figs.~\ref{fig1}(a). In the experiment, we use $D=2.5$ cm, $s=0.8$ mm and $f=10$ cm, and $\lambda_0=632$ nm. As a result, while the transverse coherence length $\sigma_c=80$ $\mu$m of the PIC source stays $z$-invariant for over 300 cm, the transverse coherence length  of the conventional source increases with $z$ as $\sigma_c=\frac{\lambda_{0} z}{s}$. The light from the source after transmitting through the object first encounters a scattering medium before getting imaged at the CCD camera. The CCD camera has $1024\times 1280$ pixels with the size of each pixel being 5 $\mu$m, the distance $d$ between the scatterer and the object is 3.5 cm, and the focal length $f$ of the imaging lens is 5 cm, which images the object with a magnification of about 3. In order to avoid the saturation of the camera, we use a neutral density (ND) filter of optical density (OD) equal to 1, placed immediately before the camera. In order to mimic objects at different  transverse planes along the direction of propagation, we keep our source at various longitudinal distances from the object. This way the imaging condition as well as the distance between the object and the scattering medium remains constant when imaging various transverse planes with different sources. Figure \ref{fig2}(b) shows the image of the object in the absence of any scattering. Figure~\ref{fig2}(c) shows images of the object obtained with the three sources at three different $z$ values and with two different scattering strength . In order to get a quantitative estimate of the image quality, we use image contrast defined as $C=(I_{\rm max}-I_{\rm min})/(I_{\rm max}+I_{\rm min})$, where $I_{\rm max}$ and $I_{\rm min}$ are the maximum and minimum intensity respectively. For calculating the contrast, we first select an area in the image, as shown by the dotted square, and then define $I_{\rm max}$ and $I_{\rm min}$ as the average pixel-intensities in the bright and dark regions within the square, respectively. We calculate the contrast of each image shown in Fig.~\ref{fig2}(c) and plot it as a function of $z$ and  for the three sources in Fig.~\ref{fig2}(d).

The results in Fig.~\ref{fig2}(c) and Fig.~\ref{fig2}(d) demonstrate how a PIC source performs imaging of different transverse planes with almost equal contrast in the presence of scattering. We find that the measured image contrast at $z=10$ cm is the same with both PIC and conventional sources, and for the two $\alpha$ values, the contrast is about 50$\%$ and 25$\%$, respectively. As $z$ is increased to $50$ cm, the contrast with the PIC source remains invariant at  50$\%$ and 25$\%$ while the contrast with the conventional source drops down to about 30$\%$ and 4$\%$, respectively, for the two $\alpha$ values. This is because $\sigma_c$ of both the sources are very similar at $z=10$ cm. However, for $z>10$ cm, $\sigma_c$ of the PIC source remains invariant while that of the conventional source increases causing the imaging contrast to decrease. As expected, the speckle effect is much more prominent for the spatially coherent source and increases with increasing scattering strengths. We note that although PIC-like sources have been earlier used in microscopy \cite{mertz2019microscopy}, here we demonstrated their usefulness in enhancing the imaging contrast at various transverse planes through a scattering medium.

\subsection{Imaging with a propagation-invariant source in reflecting configuration}

\begin{figure}[t!]
\centering
\includegraphics{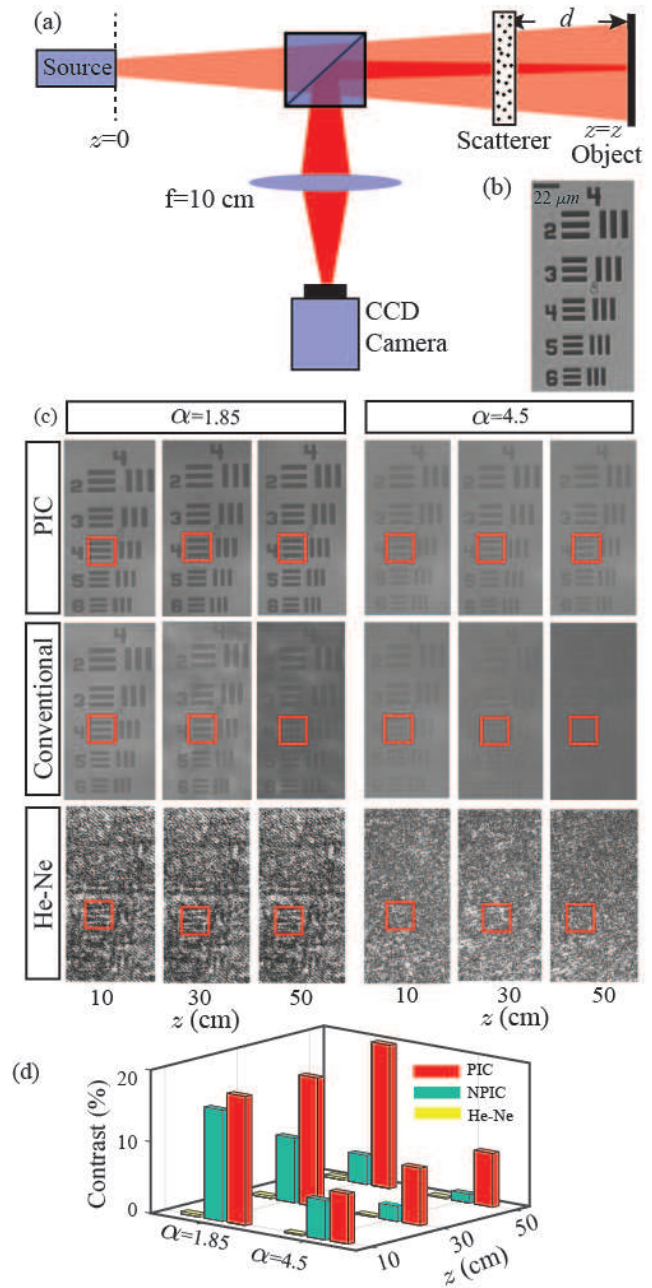}
\caption{(a) Schematic of the setup for imaging in reflecting configuration. (b) Image of the object in the absence of scattering. (c) Images of the object and (d) Imaging contrast obtained with the three different sources at different $z$ and $\alpha$ values.}\label{fig6}
\end{figure}
Although imaging in transmitting configuration is important, many real-life scenarios require imaging in reflecting configuration, in which both the source and the detector are on the same side of the object. So, next, we demonstrate the imaging capabilities of our PIC source in a reflecting configuration. Figure \ref{fig6}(a) shows the schematic diagram of the experimental setup. A source kept at $z=0$ illuminates the object at $z=z$. The light from the source first passes through a beam splitter and then after transmitting through the scatterer illuminates the object. The reflected light from the object passes through the scatterer, gets reflected by the beam splitter and then imaged at the CCD camera. The CCD camera has $1024\times 1280$ pixels with the size of each pixel being 5 $\mu$m, the distance $d$ between the scatterer and the object is 4 cm, the imaging lens of focal length $f=10$ cm images the object with a magnification of about 3. As earlier, in order to mimic the object at different transverse planes, we keep our source at various longitudinal distances from the object. Figure \ref{fig6}(b) shows the image of the object in the absence of any scattering. We use the same object as in the transmitting configuration. However, since it is a binary object with only transparent and opaque regions, the image of the object in \ref{fig6}(b) has reversed bright and dark regions as compared to the image in \ref{fig2}(b). Figure \ref{fig6}(c) shows the images of the object obtained with the three sources at three different $z$ values and with two different scattering strength . We calculate the contrast of each image shown in \ref{fig6}(c) and plot it in \ref{fig6}(d). We find that in general the results of Fig.~\ref{fig6} obtained in the reflecting configuration are qualitatively similar to those obtained in the transmitting imaging configuration. However, the contrast of the images in the reflecting configuration is lower compared to  that in the case of transmitting configuration. This is simply because in the reflecting configuration the light has to go through the scattering medium twice.

\subsection{Effect of intensity on image contrast obtained with the conventional source}
\begin{figure}[t!]
\centering
\includegraphics{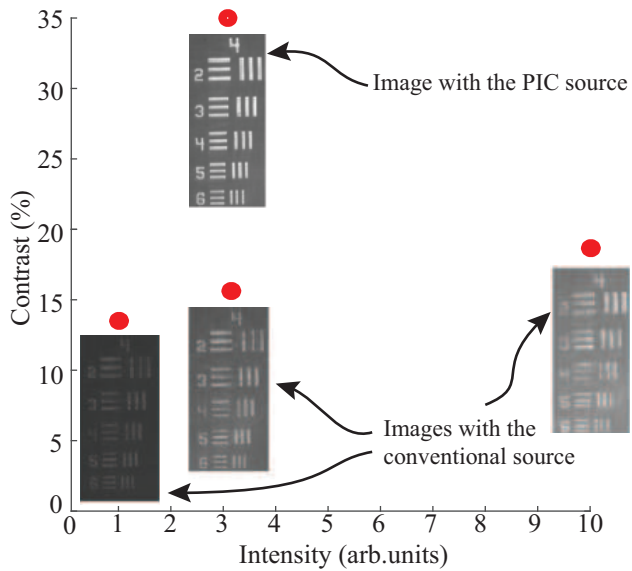}
\caption{The image and image contrast obtained with the PIC source and those with the conventional source at varying intensities. All the images were obtained at $z=30$ cm and with the scattering strength $\alpha=4.5$.}\label{fig4}
\end{figure}
In both the reflecting and transmitting imaging configurations, we find that, as $z$ increases for a given scattering strength, the image contrast as well as the illumination intensity of the images obtained with the conventional source decreases. So, a question that can arise is whether the decrease in the image contrast with increasing $z$ is due to the increase in the spatial coherence length of the source or due to the decrease in the illumination intensity. We address this question in the transmitting configuration at $z=30$ cm and $\alpha=4.5$. We record images at increased intensities of the conventional source and compare them with the image obtained with the PIC source under same experimental conditions. Figure \ref{fig4} shows one image obtained with the PIC source and the three images obtained with the conventional source at various illumination intensities. Along with the images, Fig.~\ref{fig4} also shows the corresponding image contrasts. The image obtained with the PIC source has an image contrast equal to 35$\%$. The other three images of Fig.~\ref{fig4} are obtained with the conventional source at various intensities. The first of these images is obtained under the same experimental conditions as those in the case of the PIC source, that is, with an ND filter of OD equal to 1. The illumination intensity in this case is about three times lower compared to that in the case of the PIC source. The second image is obtained with an ND filter of OD equal to 0.5 such that the illumination intensity is very close to that in the case of the PIC source. The third image is obtained with no ND filter such that the illumination intensity is increased by a factor of more than 3 compared to that in the case of the PIC source. We find that under the same experimental conditions, the image contrast with the conventional source is less than 15$\%$, compared to the 35$\%$ contrast obtained with the PIC source.  When the intensity of the conventional source is increased such that the illumination becomes comparable to that of the PIC source, the image contrast increases to only 15.6$\%$. A subsequent increase in the intensity does not improve the image contrast much further. This confirms that the decrease in the image contrast with the conventional source is indeed due to the increase in the spatial coherence length of the source and that it cannot be compensated by simply increasing the illumination intensity. Furthermore, we note that the incremental increase in the contrast as a function of the illumination intensity is due to the increased signal-to-noise ratio and  that it saturates very quickly.

\subsection{Imaging with a minimum-possible coherence source in transmitting configuration}

\begin{figure}[t!]
\includegraphics{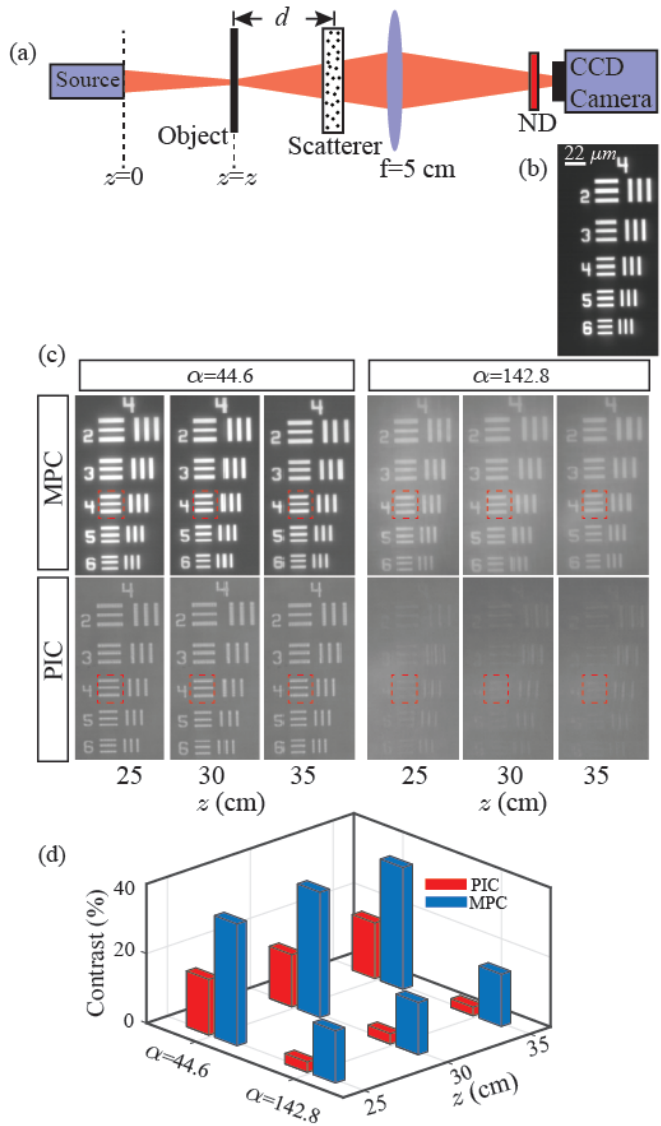}
\caption{(a) Schematic of the setup for imaging in transmitting configuration with an MPC source. (b) Image of the object in the absence of scattering. (c) Images of the object and (d) Imaging contrast  obtained with MPC and PIC sources at different $z$ and $\alpha$ values. 
}\label{fig5}
\end{figure}
Figure \ref{fig5} (a) shows the schematic experimental setup for imaging with an MPC source. We image the object kept at three different values of $z$, namely, $z=25$ cm, $z=30$ cm and $z=35$ cm. For each $z$, we choose $u$ such that the primary incoherent source gets imaged onto a plane at $z$ and the field achieves its $\sigma_{c, \rm min}$ at $z$.  The rest of our experimental setup is same as that in the case of the transmission configuration of Fig.\ref{fig2}(a). Next, in order to demonstrate enhanced imaging capabilities of our MPC source, we compare its performance with that of at PIC source under same experimental conditions. Figure~\ref{fig5}(b) shows the image of the object in the absence of any scattering. Figure~\ref{fig5}(c) shows images obtained with the two sources for three different value of $z$ and two different values of $\alpha$ . Figure~\ref{fig5}(d) shows the imaging contrast as a function of $z$ and $\alpha$. These results clearly demonstrate that the MPC source images different transverse planes with maximum possible imaging contrast. Furthermore, in the presence of a scattering medium, the MPC source provides much better imaging contrast compared not only to the conventional or coherent sources but also to the PIC source. Nevertheless, for smaller scattering strengths, a PIC source would still be preferable over an MPC source, since as opposed to the MPC source, which requires choosing a suitable $u$ for every $z$, a PIC source works with the same configuration at every $z$.

\section{Discussions and Conclusions}
     
In conclusion,  we have demonstrated that by controlling the propagation of spatial coherence it is possible to  enhance the imaging contrast at different transverse planes along the propagation direction through scattering media. Using a propagation-invariant coherence (PIC) source, we have demonstrated imaging spatially separated transverse planes without loss of contrast. Next, by making a source that has minimum-possible coherence (MPC), we have demonstrated improved imaging with maximum possible contrast. Our work can have important implications for applications that require imaging through scattering media. 

We note that in our experiments we have used scattering media of thickness ranging from $1$ mm to $6$ mm. In addition, in the reflecting configuration, we have essentially imaged an object kept between a set of two scattering media, which to some extent mimics the experimental situation in which an object is kept in a distributed scatterer. Therefore, although we have not explicitly considered the case of a distributed scatterer, which would be relevant for several realistic imaging scenarios, the results presented in this paper suggest that controlling the propagation of spatial coherence would offer similar qualitative imaging benefits even in the presence of a distributed scatterer.

\section*{Acknowledgments}

We acknowledge financial support through the research grant no. EMR/2015/001931 from the Science and Engineering Research Board (SERB), Department of Science and Technology, Government of India and through the research grant no. DST/ICPS/QuST/Theme-1/2019 from the Department of of Science and Technology, Government of India.


\begin{thebibliography}{10}

\bibitem{goodman1966apl}
J. Goodman, W. Huntley~Jr, D. Jackson, and M. Lehmann, Applied Physics Letters
  {\bf 8},  311  (1966).

\bibitem{kogelnik1968josa}
H. Kogelnik and K. Pennington, JOSA {\bf 58},  273  (1968).

\bibitem{kang2015natphot}
S. Kang, S. Jeong, W. Choi, H. Ko, T.~D. Yang, J.~H. Joo, J.-S. Lee, Y.-S. Lim,
  Q.-H. Park, and W. Choi, Nature Photonics {\bf 9},  253  (2015).

\bibitem{katz2014natphot}
O. Katz, P. Heidmann, M. Fink, and S. Gigan, Nature photonics {\bf 8},  784
  (2014).

\bibitem{bertolotti2012nature}
J. Bertolotti, E.~G. van Putten, C. Blum, A. Lagendijk, W.~L. Vos, and A.~P.
  Mosk, Nature {\bf 491},  232  (2012).

\bibitem{liba2017natcomm}
O. Liba, M.~D. Lew, E.~D. SoRelle, R. Dutta, D. Sen, D.~M. Moshfeghi, S. Chu,
  and A. de~La~Zerda, Nature communications {\bf 8},  15845  (2017).

\bibitem{goodman2007speckle}
J.~W. Goodman, {\em Speckle phenomena in optics: theory and applications}
  (Roberts and Company Publishers, Englewood, 2007).

\bibitem{ntziachristos2010natmed}
V. Ntziachristos, Nature methods {\bf 7},  603  (2010).

\bibitem{huang1991science}
D. Huang, E.~A. Swanson, C.~P. Lin, J.~S. Schuman, W.~G. Stinson, W. Chang,
  M.~R. Hee, T. Flotte, K. Gregory, C.~A. Puliafito, {\it et~al.}, Science {\bf
  254},  1178  (1991).

\bibitem{velten2012natcomm}
A. Velten, T. Willwacher, O. Gupta, A. Veeraraghavan, M.~G. Bawendi, and R.
  Raskar, Nature communications {\bf 3},  745  (2012).

\bibitem{leith1992josaa}
E. Leith, C. Chen, H. Chen, Y. Chen, D. Dilworth, J. Lopez, J. Rudd, P.-C. Sun,
  J. Valdmanis, and G. Vossler, JOSA A {\bf 9},  1148  (1992).

\bibitem{harm2014optexp}
W. Harm, C. Roider, A. Jesacher, S. Bernet, and M. Ritsch-Marte, Optics express
  {\bf 22},  22146  (2014).

\bibitem{vellekoop2007optlett}
I.~M. Vellekoop and A. Mosk, Optics letters {\bf 32},  2309  (2007).

\bibitem{mosk2012natphot}
A.~P. Mosk, A. Lagendijk, G. Lerosey, and M. Fink, Nature photonics {\bf 6},
  283  (2012).

\bibitem{ohtsuka1980optcomm}
Y. Ohtsuka, Y. Nozoe, and Y. Imai, Optics Communications {\bf 35},  157
  (1980).

\bibitem{chen2014pra}
Y. Chen, F. Wang, L. Liu, C. Zhao, Y. Cai, and O. Korotkova, Phys. Rev. A {\bf
  89},  013801  (2014).

\bibitem{wang2014optexp}
F. Wang, C. Liang, Y. Yuan, and Y. Cai, Optics express {\bf 22},  23456
  (2014).

\bibitem{basu2014optexp}
S. Basu, M.~W. Hyde, X. Xiao, D.~G. Voelz, and O. Korotkova, Optics express
  {\bf 22},  31691  (2014).

\bibitem{ostrovsky2009optexp}
A.~S. Ostrovsky, G. Mart{\'\i}nez-Niconoff, V. Arriz{\'o}n, P.
  Mart{\'\i}nez-Vara, M.~A. Olvera-Santamar{\'\i}a, and C. Rickenstorff-Parrao,
  Optics express {\bf 17},  5257  (2009).

\bibitem{redding2012natphot}
B. Redding, M.~A. Choma, and H. Cao, Nature photonics {\bf 6},  355  (2012).

\bibitem{redding2015pnas}
B. Redding, A. Cerjan, X. Huang, M.~L. Lee, A.~D. Stone, M.~A. Choma, and H.
  Cao, Proceedings of the National Academy of Sciences {\bf 112},  1304
  (2015).

\bibitem{carter1977josa}
W. Carter and E. Wolf, JOSA {\bf 67},  785  (1977).

\bibitem{tziraki2000apb}
M. Tziraki, R. Jones, P. French, M. Melloch, and D. Nolte, Applied Physics B:
  Lasers and Optics {\bf 70},  151  (2000).

\bibitem{redding2015optlett}
B. Redding, P. Ahmadi, V. Mokan, M. Seifert, M.~A. Choma, and H. Cao, Optics
  letters {\bf 40},  4607  (2015).

\bibitem{knitter2016optica}
S. Knitter, C. Liu, B. Redding, M.~K. Khokha, M.~A. Choma, and H. Cao, Optica
  {\bf 3},  403  (2016).

\bibitem{zheng2016optexp}
Y. Zheng, J. Si, W. Tan, Y.~H. Ren, J. Tong, and X. Hou, Optics express {\bf
  24},  26338  (2016).

\bibitem{farrokhi2017scirpt}
H. Farrokhi, T.~M. Rohith, J. Boonruangkan, S. Han, H. Kim, S.-W. Kim, and
  Y.-J. Kim, Scientific Reports {\bf 7},  15318  (2017).

\bibitem{goodman2015stat}
Joseph W Goodman, {\em Statistical optics} (John Wiley \& Sons, Hoboken, 2015) 

\bibitem{aarav2017pra}
S. Aarav, A. Bhattacharjee, H. Wanare, and A.~K. Jha, Physical Review A {\bf
  96},  033815  (2017).

\bibitem{mandel1995coherence}
L. Mandel and E. Wolf, {\em Optical Coherence and Quantum Optics} (Cambridge
  university press, New York, 1995).

\bibitem{wang2012biomedical}
L.~V. Wang and H.-i. Wu, {\em Biomedical optics: principles and imaging} (John
  Wiley \& Sons, New York, 2012).

\bibitem{mertz2019microscopy}
J. Mertz, {\em Introduction to optical microscopy} (Cambridge University Press,
  New York, 2019).

\end{thebibliography}

\end{document}